\documentclass
[preprint,pre,nofootinbib,tightenlines,showpacs,notitlepage]{revtex4}%
\usepackage{amsfonts}
\usepackage{amsmath}
\usepackage{amssymb}
\usepackage{graphicx}
\usepackage{acronym}%
\begin{document}
\title{Towards a general theory of nonlinear flow-distributed oscillations}
\author{Patrick N. McGraw and Michael Menzinger}
\affiliation{Department of Chemistry, University of Toronto, Toronto, Ontario, Canada M5S 3H6}

\begin{abstract}
We outline a general theory for the analysis of flow-distributed standing and
travelling wave patterns in one-dimensional, open plug-flows of oscillatory
chemical media. \ We treat both the amplitude and phase dynamics of small and
large-amplitude waves, considering both travelling and stationary waves on an
equal footing and emphasizing features that are generic to a variety of
kinetic models. \ We begin with a linear stability analysis for constant and
periodic boundary forcing, drawing attention to the implications for systems
far from a Hopf bifurcation. Among other results, we show that for systems far
from a Hopf bifurcation, the first absolutely unstable mode may be a
stationary wave mode. \ We then introduce a non-linear formalism for studying
both travelling and stationary waves and show that the wave forms and their
amplitudes depend on a single reduced transport parameter. \ Our formalism
sheds light on cases where neither the linearized analyis nor the kinematic
theory of phase waves give an adequate description, and it can be applied to
study some of the more complex types of bifurcations (Canards,
period-doublings, etc.) in open flow systems. \ 

\end{abstract}
\maketitle

\section{Introduction}

Among the variety of known mechanisms of spontaneous pattern formation in
spatially extended chemical systems, the flow distributed oscillation (FDO)
mechanism \cite{Kuznetsov}-\cite{Satnoianu2001} differs from the Turing
\cite{Turing} and related mechanisms\cite{Muratov} and from the differential
flow instability\cite{DIFICI}\cite{Toth2001} in that it does not require any
differential transport of the reacting species. \ The FDO mechanism, which was
predicted\cite{Kuznetsov}\cite{Andresen} physically interpreted and
experimentally verified\cite{Kaern99a}-\cite{Kaern2002}, operates in open flow
systems when the chemical kinetics is intrinsically oscillatory and the inflow
boundary condition plays an essential role. \ It is potentially relevant to
biological pattern formation involving an axial growth, since an open flow
system is related by a coordinate transformation to a linearly growing medium
such as a plant stem or animal embryo.\cite{KMH2000a}-\cite{Movebound}

The governing equations of the one-dimensional open reactive flow studied are
the reaction-diffusion-advection \ (RDA) equation
\begin{equation}
\frac{\partial\mathbf{U}}{\partial t}=\mathbf{f}(\mathbf{U;}C)-v\frac
{\partial\mathbf{U}}{\partial x}+D\frac{\partial^{2}\mathbf{U}}{\partial
x^{2}}. \label{RDAeq}%
\end{equation}
together with the boundary condition \ $\mathbf{U}(0,t)$\ \ at the inflow.
\ Here\ $\mathbf{U}(x,t)$\ is an $N$-dimensional vector of dynamical variables
(concentrations of species), \ $\mathbf{f}(\mathbf{U;}C)$ is the vector-valued
rate function which depends on one or more control parameters $C$, $v>0$ is
the flow velocity and $D$ is the diffusion coefficient. \ In general $D$ and
$v$ can be matrices, allowing for differential transport, but here we focus on
the case without differential transport, so that $v$ and $D$ are real
scalars.\ We take the length of the reactor to be $L$ and impose no-flux
boundary conditions at the outflow: \ $\left.  \partial U/\partial
x\right\vert _{x=L}=0$. \ We are interested in the case where $\mathbf{f}%
(\mathbf{U;}C)$ has a stable limit cycle and at least one unstable fixed
point. \ 

Insight into the physical mechanism of flow-distributed waves can be gained by
considering the \textit{kinematic limit} \cite{Reply} of zero diffusion. \ In
this case equation (\ref{RDAeq}) can be written as%
\begin{equation}
\frac{d\mathbf{U}}{dt}=\mathbf{f}(\mathbf{U};C), \label{kinematic}%
\end{equation}
where we have introduced the advective derivative $d/dt\equiv\partial/\partial
t+v\partial/\partial x$. Then the evolution of an advected fluid element is
the same as that of the well-mixed system: each fluid element is an
independent oscillator or batch reactor, and the inflow boundary condition
serves to establish the phase of \ these oscillators.\cite{Kaern99a} \ If the
boundary condition is constant, for example, then all of the oscillators enter
the reactor with the same initial phase and stationary waves result from the
recurrence of the same phase at equally spaced downstream positions. \ On the
other hand, an oscillating boundary condition leads to upstream or downstream
travelling waves, which have also been verified experimentally.\cite{KMH2000a}%
-\cite{Satnoianu2001} \ While the kinematic limit is helpful for understanding
the essential physics, \ there are significant deviations from its predictions
when diffusion becomes significant $D\neq0.$ These deviations affect the
wavelengths\cite{Comments} \ as well as the amplitudes and shapes of flow
distributed waves. \ The path in phase space of a volume element moving
through the reactor may differ from that of the well-stirred system with the
same initial conditions.\cite{Kaern2000} \ Sufficiently strong diffusion can
extinguish the waves due to the diffusive mixing of adjacent fluid elements
which are oscillating out of phase. \ 

An alternative aproach to this kinematic or phase dynamics picture is the
original \cite{Kuznetsov}\cite{Andresen} linear stability analysis which views
the flow-distributed oscillations as arising from growing perturbations of an
unstable uniform stationary state $\mathbf{U}(x,t)=\mathbf{U}_{0}$ where
$\mathbf{U}_{0}$ is an unstable fixed point of the underlying kinetics
governed by $\mathbf{f}(\mathbf{U;}C)$. \ This linearized analysis is useful
for predicting which small perturbations initially grow and which do not and
it has been used in a variety of systems to determine thresholds for the
formation of wave patterns (for example, the bifurcation from growing to
evanescent stationary waves.) \ However, the linearized analysis is not
sufficient to determine the behavior of the waves when their amplitude becomes
large. \ As was pointed out in ref. \cite{Bamforth2001}, the linearized
analysis is most useful when the kinetic system is not far from a
supercritical Hopf bifurcation. In this case, it can give a good approximation
to the wavelength of the fully developed nonlinear waves, although by itself
it is inadequate to predict their amplitude or shape. \ Chemical oscillators,
however, have interesting dynamical regimes far from a Hopf bifurcation,
showing markedly non-sinusoidal or relaxation oscillations. \ In such cases,
the fixed point enclosed by the limit cycle may be an unstable node rather
than a focus, in which case a linearized analysis near the fixed point does
not reveal the intrinsic oscillatory behavior at all. \ The behavior of large
perturbations can then be radically different from that of small ones. \ 

Our goal in this work is to develop a general formalism that describes both
stationary and travelling flow-distributed waves including their nonlinear
behavior and deviations from the kinematic limit. \ This formalism provides a
unifying framework for understanding previous results and it can be applied to
arbitrary kinetic models. \ In contrast to previous studies of particular
kinetic models\cite{Kuznetsov}\cite{Andresen}\cite{Bamforth2001}%
\cite{Bamforth2002}, our emphasis is on generic phenomena. In particular we
hope to shed light on systems with relaxation or other non-sinusoidal
oscillations, where the unstable fixed point is a node rather than a focus.
\ \ \ In Section II we discuss the linear stability analysis of a fixed point
with an emphasis on generic features of the two types of fixed points,
unstable foci and unstable nodes. \ We derive a general dispersion relation
for small-amplitude disturbances and use it to analyze the response to
time-dependent perturbations imposed at the inflow. \ We show that a band of
frequencies is spatially amplified, and that this band becomes narrower and
sharper with increasing diffusion. \ We derive general expressions for the
thresholds separating absolute from convective instability as well as the
threshold for extinction of stationary waves and show that the latter
threshold disappears in the case of an unstable node. \ \ If the system is far
from a Hopf bifurcation, we show that, in contrast to previous examples, it is
possible for sustained stationary waves to arise through an absolute
instability from a transient perturbation. \ In Section III we use a
travelling wave ansatz to introduce a reduced one-dimensional ordinary
differential equation (ODE) that describes the fully nonlinear stationary and
travelling waves. \ Numerical solutions of this equation are easier to obtain
than those of the partial differential equation (\ref{RDAeq}) and they allow
us to obtain the \textit{nonlinear dispersion relation}\ for the fully
developed, large amplitude waves. A key result is that the amplitude and shape
of the wave depend only on the reduced transport parameter $\Gamma\equiv
D/(v-c)^{2}$ where $c$ is the phase velocity of the wave. \ It is $\Gamma$
that governs the strength of deviations from the kinematic limit. \ We show
that in the case of relaxation oscillations these deviations are qualitatively
different from those in the quasi-sinusoidal case, \ and we provide a physical
interpretation of the dependence on $\Gamma$. \ We illustrate the application
of our techniques by some numerical results. \ \ Finally, in Section IV we
summarize our conclusions and describe the prospects for applying our
techniques to more complex types of bifurcations (Canards, period doublings,
etc.) \ in open flow systems. \ 

Throughout the paper we use as an illustrative model the van der Pol or
FitzHugh-Nagumo (FN) oscillator\cite{Muratov}\cite{FN1Dawson}\cite{FN2Hagberg}%
\begin{align}
\frac{dX}{dt}  &  =e(X-X^{3}-Y)\label{FHN}\\
\frac{dY}{dt}  &  =-Y+10X,\nonumber
\end{align}
which is described in more detail in Appendix A. \ The FN model exhibits the
generic features we are interested in: \ it provides examples of
quasi-sinusoidal oscillations changing to relaxation oscillations and an
unstable focus changing to an unstable node as $e$ is increased. \ Similar
qualitative features occur in the Brusselator, Oregonator and other kinetic
models. Appendix B describes the numerical approaches used and challenges
encountered in solving the 1-D ODE described in Section III. \ 

\section{Generic linearized analysis near a fixed point}

In this section we pursue the approach which views flow-distributed waves as
arising from instabilities of the spatially uniform solution \ $\mathbf{U}%
(x,t)=\mathbf{U}_{0}$ of equation (\ref{RDAeq}). \ The systems of interest
possess an unstable fixed point and a stable limit cycle. \ The instability of
the uniform state may be either convective or absolute.\cite{Deissler}%
\cite{Whitham}\cite{Huerre} \ We now consider small perturbations
$\mathbf{u}(x,t)$ \
\begin{equation}
\mathbf{U}(x,t)=\mathbf{U}_{0}+\mathbf{u}(x,t). \label{smallpert}%
\end{equation}
of the homogeneous fixed point $\mathbf{U}_{0}.$\ In the linearized
approximation, $\mathbf{u}(x,t)$ obeys%
\begin{equation}
\frac{\partial\mathbf{u}}{\partial t}=\mathbb{J}(\mathbf{U}_{0})\mathbf{u}%
-v\frac{\partial\mathbf{u}}{\partial x}+D\frac{\partial^{2}\mathbf{u}%
}{\partial x^{2}} \label{linearized}%
\end{equation}
where $\mathbb{J}$ is the Jacobian matrix $\partial\mathbf{f}(\mathbf{U}%
)/\partial\mathbf{U}.$\ Let us denote the eigenvectors and eigenvalues of the
Jacobian by $\mathbf{\xi}_{j}$ and $\lambda_{j}$ respectively ($j\in\{1..N\}$)
and expand the perturbation $\mathbf{u}(x,t)$ in the eigenbasis as
\begin{equation}
\mathbf{u}(x,t)=\sum_{j=1}^{N}u_{j}(x,t)\mathbf{\xi}_{j}. \label{expansion}%
\end{equation}
Substitution into the linearized equation gives a separate equation for each
component:%
\begin{equation}
\frac{\partial u_{j}}{\partial t}=\lambda_{j}u_{j}-v\frac{\partial u_{j}%
}{\partial x}+D\frac{\partial^{2}u_{j}}{\partial x^{2}}. \label{componenteq}%
\end{equation}

We are interested in the dynamics along unstable eigenvectors, \ i.e., ones
for which the real part of $\lambda_{j}$ is positive. \ $\lambda_{j}$ may be
one of a pair of complex conjugate eigenvalues $\lambda_{\pm}=\alpha\pm
i\beta$ with $\alpha$ positive, \ in which case $\mathbf{U}_{0}$ is an
unstable focus,\ or it may be purely real and positive, as for an unstable
node. \ In general, the former case occurs in the neighborhood of a
supercritical Hopf bifurcation, while the latter \ may occur farther from the
bifurcation. \ The component equation associated with the eigenvector $\xi
_{j}$ can be written as%
\begin{equation}
\frac{\partial u_{j}}{\partial t}=(\alpha+i\beta)u_{j}-v\frac{\partial u_{j}%
}{\partial x}+D\frac{\partial^{2}u_{j}}{\partial x^{2}}. \label{pluscomponent}%
\end{equation}
Without loss of generality we can assume that $\beta\geqslant0$. \ The case of
a purely real eigenvalue is included by setting $\beta=0$. \ We seek complex
exponential solutions of this equation of the form
\begin{equation}
u(x,t)=A\exp(i\omega t+ikx). \label{complexexp}%
\end{equation}
(From here on, the subscript $j$ is suppressed.) \ Both $\omega$ and $k$ can
be complex: $\omega=\omega_{r}+i\omega_{i}$ and $k=k_{r}+ik_{i}$. \ Our sign
conventions are such that positive values of $\omega_{i}$ or $k_{i}$
correspond to solutions that damp with time or downstream distance
respectively. \ Substitution of the complex exponential ansatz
(\ref{complexexp}) into equation (\ref{pluscomponent}) gives the dispersion
relation%
\begin{equation}
i\omega=\alpha+i\beta-ivk-Dk^{2}. \label{dispersion}%
\end{equation}
\ We now consider modes with $\omega_{i}=0$, which represent the response to a
sinusoidal forcing at the boundary with frequency $\omega_{r}.$ With this
restriction to real frequencies, the real and imaginary components of the
dispersion relation (\ref{dispersion}) read
\begin{equation}
vk_{i}+D(k_{i}^{2}-k_{r}^{2})=-\alpha, \label{dispreal}%
\end{equation}%
\begin{equation}
vk_{r}+2Dk_{r}k_{i}=\beta-\omega_{r}. \label{dispim}%
\end{equation}
\ If $\omega_{r}=\beta$ (forcing at the natural oscillation frequency), then
eq. (\ref{dispim}) can be satisfied by setting $k_{r}=0.$ Eq. (\ref{dispreal})
is then reduced to a quadratic equation in $k_{i}$, which always has a
negative (spatially growing) solution provided $D/v^{2}<1/4\alpha$. \ (As we
will discuss below, $D/v^{2}=1/4\alpha$ marks the threshold between convective
and absolute instability of the fixed point. \ Beyond this threshold the
dispersion relation cannot be solved with a purely real $\omega$.) \
$>$%
$>$%
From this we learn that a perturbation with $\omega_{r}=\beta$ always yields a
growing mode (the Hopf mode) in which the whole reactor oscillates in phase.
\ In fact, we shall see that $\beta$ is the midpoint of a band of amplified
frequencies. \ The edges of the amplified band can be derived by setting
$k_{i}=0$ in eqs. (\ref{dispreal}) and (\ref{dispim}) and solving for
$\omega_{r}$. \ The result is that sinusoidal perturbations are spatially
amplified if
\begin{equation}
\beta-\sqrt{\frac{\alpha v^{2}}{D}}<\omega_{r}<\beta+\sqrt{\frac{\alpha v^{2}%
}{D}} \label{amprange}%
\end{equation}
and they are damped otherwise. \ If
\begin{equation}
\frac{D}{v^{2}}>\frac{\alpha}{\beta^{2}} \label{statthreshhold}%
\end{equation}
then $\omega_{r}=0$ falls outside the range of amplified frequencies and the
stationary waves decay. \ 

\subsection{Threshold between Absolute and Convective Instability}

An important distinction in the dynamics of open flow systems is the one
between absolute and convective instability. \ \ If a state is
\textit{convectively unstable}, then perturbations of the state grow in a
reference frame moving with the flow velocity, but for any fixed position in
the laboratory frame the disturbance eventually decays. \ The effects of a
perturbation of finite duration are eventually washed downstream and out of
the system and the reactor returns to its initial unperturbed state. \ In this
case, the initial conditions at $t=0$ do not influence the asymptotic
late-time behavior, which is instead controlled by the upstream boundary
condition. \ If a state is \textit{absolutely unstable}, on the other hand,
then localized perturbations grow with time at a fixed position in the
laboratory frame, propagating upstream as well as downstream. This means that
perturbations applied at $t=0$ continue to affect the state at late times.
\ From an experimental point of view, \ the continuing effect of noise in the
initial conditions makes an absolutely system more complicated. \ Since the
disturbances propagate in both directions, the system resembles a pure
reaction-diffusion system where the flow plays a subordinate role. The
critical slowing-down near the convective/absolute instability threshold adds
to the experimental difficulty. \ A convectively unstable system can be
controlled more readily by manipulating the inflow boundary condition. \ 

The distinction between the two cases can be expressed in terms of the
\emph{group velocity} of propagating disturbances.\cite{Andresen}%
\cite{Bamforth2000}\cite{Whitham}\cite{Huerre} If there are modes with
positive growth rate and upstream (negative) group velocity then the
instability is absolute --- otherwise, all growing disturbances are washed
downstream.\ \ The threshold between the two cases occurs when a mode with
zero group velocity \ $d\omega/dk=0$ has exactly zero growth with respect to
time, $\omega_{i}=0$. \ This represents a persistent disturbance that is not
washed downstream. \ From the dispersion relation (\ref{dispersion}) the
condition of zero group velocity is
\[
0=\frac{d\omega}{dk}=-iv-2Dk=0
\]
or%
\begin{equation}
k=\frac{-iv}{2D}, \label{zerogroup}%
\end{equation}
giving a purely imaginary wave number which can be substituted into both
components of the dispersion relation (\ref{dispreal}) and (\ref{dispim}).
\ The threshold for absolute instability occurs when the temporal growth rate
$-\omega_{i}$ is exactly zero for this zero-group-velocity mode, so from the
real component (\ref{dispreal}) we get%
\[
0=-\omega_{i}=\alpha-\frac{v^{2}}{2D}+D(\frac{v}{2D})^{2}=\alpha-\frac{v^{2}%
}{4D}%
\]
or%
\begin{equation}
\frac{D}{v^{2}}=\frac{1}{4\alpha} \label{AIthreshold}%
\end{equation}
\ \ From the imaginary component (\ref{dispim}) we get that $\omega_{r}=\beta
$, which tells us that the first mode to go absolutely unstable is the Hopf
mode. \ 

Note that eq. (\ref{AIthreshold}) \ is the same threshold mentioned above in
discussing the solutions of (\ref{dispreal}) and (\ref{dispim}). \ It is the
threshold beyond which the most unstable mode grows with time as well as space
and therefore its frequency cannot be purely real.

\subsection{Universal solutions of the dispersion relation}

\begin{figure}
[ptb]
\begin{center}
\includegraphics[
height=3.1834in,
width=4.843in
]%
{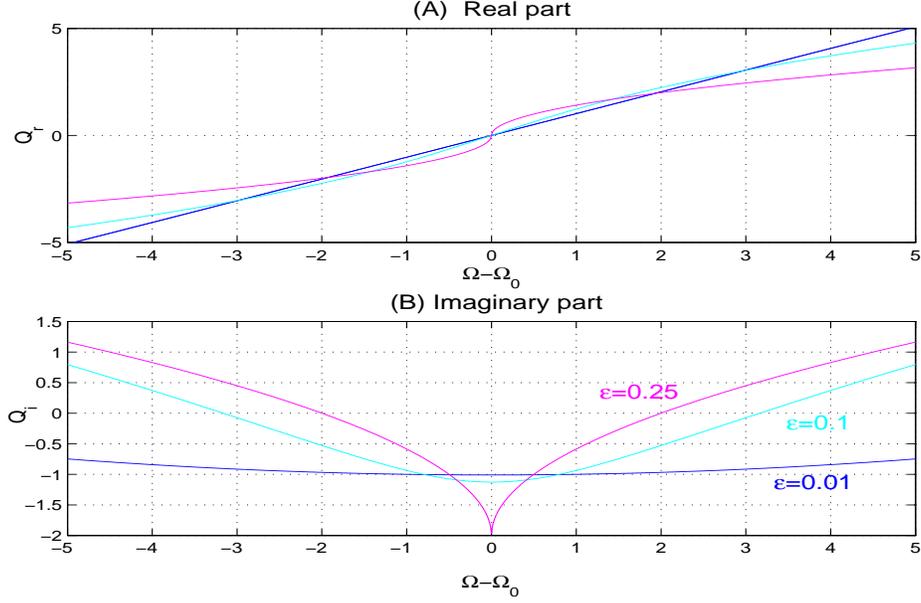}%
\caption{General solutions to the dispersion relation for small-amplitude
oscillations near an unstable fixed point. \ All frequencies are scaled
relative to the inverse growth time scale $\alpha.$ \ Recall that negative
values of the imaginary wavenumber $Q_{i}$ represent growing modes. \ }%
\label{relaxdisp}%
\end{center}
\end{figure}

To illustrate the above results generically, it is helpful to express the
dispersion relation in terms of dimensionless variables. \ This requires a
choice of a characteristic time scale. \ It is convenient to use the growth
time scale $\tau=1/\alpha$ and introduce the variables $\Omega\equiv
\omega/\alpha$, $Q=Q_{r}+iQ_{i}\equiv vk/\alpha$, $\varepsilon\equiv
D\alpha/v^{2}$ and $\Omega_{0}\equiv\beta/\alpha$. \ For real $\Omega$ the two
components (\ref{dispreal}) and (\ref{dispim}) of the dispersion relation can
be rewritten as%
\begin{equation}
-Q_{i}+\varepsilon(Q_{r}^{2}-Q_{i}^{2})=1, \label{scaleddispreal}%
\end{equation}%
\begin{equation}
Q_{r}+2\varepsilon Q_{r}Q_{i}=\Omega-\Omega_{0}. \label{scaleddispim}%
\end{equation}
Numerical solutions for $Q_{r}$ and $Q_{i}$ as functions of $\Omega-\Omega
_{0}$ are plotted in figure \ref{relaxdisp}\ for several values of
$\varepsilon$. \ The family of curves illustrates the qualitative features of
the response to sinusoidal forcing: the band of amplified frequencies becomes
becomes sharper and narrower with increasing $\varepsilon,$ and the plot of
the wavenumber $Q_{r}$ versus $\Omega$ becomes more strongly nonlinear.
\ \ The maximum growth rate always occurs at $\Omega=\Omega_{0}$. The onset of
absolute instability occurs when $\varepsilon=1/4$ \ and is marked by the
appearance of a cusp in $Q_{i}$ and a vertical jump in $Q_{r}$ at
$\Omega=\Omega_{0}$. \ In the kinematic limit $\varepsilon\rightarrow0$ all
frequencies are amplified equally and $Q_{r}$ is a linear function of $\Omega
$. \

\subsection{Comments on the unstable node case}

In the preceding paragraphs we have derived results for the linear stability
analysis of perturbations along a single eigendirection near a fixed point of
the dynamics. \ The results are formally independent of whether the fixed
point is a focus or a node. \ The latter case can be described simply by
considering $\beta=0$. \ \ However, there are important qualitative
differences between the two cases. \ In the case of an unstable focus, the
eigenvalues are complex conjugates and trajectories spiral away from the fixed
point. \ There is a single growth rate $\alpha$. \ For a node, on the other
hand, there is no intrinsic oscillatory behavior near the fixed point and any
oscillation must be imposed externally or arise from nonlinear effects at
larger amplitudes. \ There are generally two eigendirections associated with
distinct growth rates $\alpha_{j}$ and therefore different dispersion
relations and instability thresholds for perturbations along the two
eigendirections. \ Perturbations grow most rapidly along the most unstable
eigendirection and therefore this eigendirection is physically the most
important. \ 

\begin{figure}
[ptb]
\begin{center}
\includegraphics[
height=3.6097in,
width=5.3428in
]%
{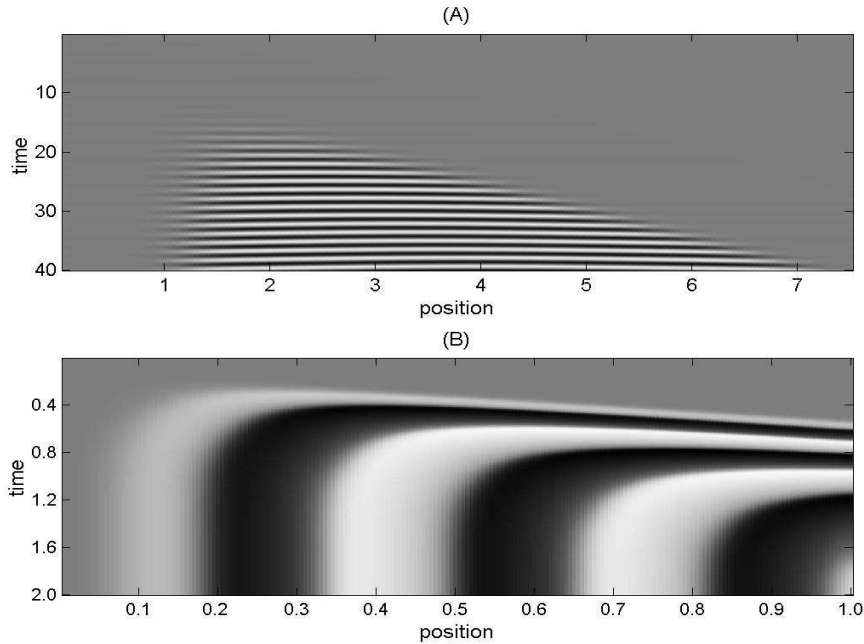}%
\caption{Onset of absolute instability in the FN system. \ The gray levels are
proportional to $X(x,t)$. The inflow boundary conditions are $X(0,t)=-0.001,$
$0<t<T$, \ $X(0,t)=0$ otherwise, where $T=2$ in case (A), $0.1$ in case (B).
\ If the system were convectively unstable, the effects of the perturbation
would be washed out of the system, but instead both systems are just above the
threshold of absolute instability and the effects of the disturbance persist.
\ The pulse perturbation has a fourier spectrum of frequencies, of which the
most unstable grows most rapidly. \ In case (A) ($e=2$, $v=0.1,$ $D=0.0052$)
\ the fixed point is a focus and the first absolutely unstable mode is
oscillatory. \ In case (B) ($e=50,$ $v=1,$ $D=0.01$) the fixed point is a node
and the first absolutely unstable mode is a stationary wave. \ Although the
linearized theory correctly predicts that the persistent mode is a stationary
wave, it does not predict the finite wavelength, nor does it adequately
describe the downstream front of the disturbance. \ }%
\label{AIcombined}%
\end{center}
\end{figure}

Another difference between the two cases is that for a focus ($\beta\neq0$)
\ the most unstable mode is always the Hopf mode which is a uniform
oscillation at frequency $\beta$. \ The band of amplified perturbations is
centered on this nonzero frequency, which is also the first mode to become
absolutely unstable. When $\beta=0,$ on the other hand, the most unstable mode
(and the first to become absolutely unstable) is a stationary mode with
$\omega_{r}=0.$ \ The linearized dispersion relation predicts a real component
of the wavenumber $k_{r}=0$ for this mode. \ However, when the amplitude
grows, the nonlinear terms grow rapidly with the result that large-amplitude
oscillation and stationary waves of finite wavelength take over. \ The
different nature of the absolute instability threshold between the two cases
is illustrated with numerical results in figure \ref{AIcombined}%
.\footnote{Numerical solutions of the RDA equation were obtained using a
simple first-order discretization. \ The time and space grids were adjusted
according to the characteristic time and space scales of the system being
studied. \ For a sufficiently fine grid, the results were verified to be
insensitive to the grid size. \ } \ The emergence of sustained stationary
waves due to absolute instability from a time-limited perturbation (fig.
\ref{AIcombined}b) is in sharp contrast to previous
examples\cite{Bamforth2000}\cite{Bamforth2001} \ in which the first absolutely
unstable mode was time-dependent (as in fig. \ref{AIcombined}a) and stationary
waves in the absolutely unstable regime only occured in a thin boundary
region, if at all, and only in response to a steady boundary condition held
away from the fixed point. \ \

\subsection{Bifurcation Loci in Control Parameter Space}

In a reactive flow experiment, it is typically the flow velocity which is
under the experimenter's control while the diffusion constant is fixed. \ One
focus of previous studies has therefore been the determination of critical
flow velocities at which the FDO\ patterns change qualitatively. \ Two
important thresholds are the threshold for the formation of undamped
stationary patterns $v_{T}$ and the threshold between absolute and convective
instability, $v_{c},$ using the notation of ref. \cite{Andresen} .\ \ General
expressions for these threshold velocities are readily obtained from
(\ref{statthreshhold}) and (\ref{AIthreshold}):%
\begin{align}
v_{T}(C)  &  =\frac{\beta(C)}{\sqrt{\alpha(C)}}\frac{1}{\sqrt{D}}\label{VT}\\
v_{c}(C)  &  =2\sqrt{\alpha(C)}\frac{1}{\sqrt{D}}, \label{Vc}%
\end{align}
where the dependence on kinetic control parameters $C$ has been made explicit.
\ These thresholds were plotted as functions of a control parameter for
particular systems in refs. \ \cite{Andresen},\cite{Bamforth2000} and
\cite{Bamforth2001}. \ We can now understand some generic features of the
shapes of these bifurcation loci observed in those works. \ Near a
supercritical Hopf bifurcation, \ $v_{T}$ diverges and $v_{c}\rightarrow0$
because $\alpha\rightarrow0$. \ In a range above the Hopf bifurcation, then,
$v_{T}>v_{c}$ and there are three dynamical regimes as the flow velocity is
varied. \ At high velocities, there are sustained stationary waves which can
be excited by fixed boundary conditions. \ For $v_{c}<v<v_{T}$, the stationary
waves are \textit{evanescent} and penetrate only a limited distance into the
medium. \ After any initial transients\footnote{These transients can
themselves have complex and interesting structures which we do not consider
here; see, e.g., ref. \cite{Bamforth2002}.} are convectively washed out of the
system, the remainder of the system returns to the fixed point. \ Thus there
is a range of $v$ values for which the unstable fixed point is effectively
stabilized in the presence of constant boundary conditions. \ (However,
oscillatory perturbations in the amplified band still result in undamped
travelling waves.) \ \ As a control parameter is tuned farther away from the
Hopf bifurcation, $\alpha(C)$ generally increases, and the two curves
$v_{T}(C)$ and $v_{c}(C)$ may cross (see, for example, figure 1 of ref.
\cite{Andresen}) so that for some values of $C$, $v_{T}<v_{c}$. \ In this case
there is no intermediate regime of evanescent waves followed by return to the
fixed point as described above. \ Instead, as the velocity is lowered, the
absolute instability threshold is reached first. \ Below $v_{c}$, \ the
effects of initial conditions at $t=0$ are not washed out of the system\ and
the late-time behavior may not be controllable by the boundary condition.
\ The most unstable modes dominate, and stationary waves may occur only in a
boundary layer with constant boundary conditions. \ \ It is possible that the
crossing of the two thresholds ($v_{T}<v_{c}$) accounts in part for the
failure to observe evanescent stationary waves in experiments.\cite{Kaern99a}%
\cite{Bamforth2001}\ In some cases, \ as $C$ is tuned still further from a
Hopf bifurcation, $\beta(C)$ may decrease to zero, and the threshold $v_{T}$
vanishes as the fixed point becomes an unstable node. \ In this case,
stationary waves may be triggered by an absolute instability as mentioned
above. \ 

\section{Nonlinear, Large Amplitude Waves}

In this section we derive a reduced ordinary differential equation (ODE) that
describes both stationary and travelling wave solutions of the RDA equation
(\ref{RDAeq}) and that applies to situations where neither the linear
stability analysis nor the kinematic limit give an adequate description. \ We
show that the amplitude and wave form depend on a single reduced transport
parameter which\ characterizes the degree of departure from the kinematic
limit. The application of our formalism is then illustrated by some numerical
examples. \ As in the linearized analysis, we find that the two cases near and
far from a Hopf bifurcation give qualitatively different wave behavior.

We make the \emph{ansatz} $\mathbf{u}(x,t)=\mathbf{u}_{c}(x-ct)$ which is
analogous to the D'Alembert solution of the wave equation. \ It represents a
generic wave (not necessarily periodic) travelling downstream with velocity
$c$ and depending only on the combination $\zeta\equiv x-ct$. \ A growing or
decaying travelling wave is not strictly described by this \emph{ansatz}
unless $c=0$ (since a spatially changing amplitude implies a dependence on $x$
and not purely on $\zeta)$ but we expect that it describes the asymptotic
behavior of such a wave when the amplitude saturates. \ Substituting this into
the reaction-diffusion-advection equation (\ref{RDAeq}) \ gives a
one-dimensional ODE: \
\begin{equation}
0=\mathbf{f}(\mathbf{u})-(v-c)\frac{\partial\mathbf{u}}{\partial\zeta}%
+D\frac{\partial^{2}\mathbf{u}}{\partial\zeta^{2}}. \label{travelODE}%
\end{equation}
With a change of variable $\zeta^{\prime}\equiv\zeta/($ $v-c)$ \ \ we obtain
\begin{equation}
0=\mathbf{f}(\mathbf{u})-\frac{d\mathbf{u}}{d\zeta^{\prime}}+\Gamma\frac
{d^{2}\mathbf{u}}{d\zeta^{\prime2}}, \label{travelODEscaled}%
\end{equation}
where $\Gamma\equiv D/(v-c)^{2}$. \ \ \ $\Gamma$ represents the effective
strength of the diffusion term for a given wave. \ If (\ref{travelODEscaled})
has a periodic solution $\mathbf{u}(\zeta^{\prime})=\mathbf{u}(\zeta^{\prime
}+\Lambda)$, then the period $\Lambda$ in terms of $\zeta^{\prime}$ is related
to the frequency and wave number of the corresponding travelling wave by%
\begin{equation}
\omega=ck=\frac{c}{c-v}\frac{2\pi}{\Lambda(\Gamma)}, \label{biglambda}%
\end{equation}
where we have made explicit the dependence of $\Lambda$ on $\Gamma$.

Some limiting cases are instructive and help to furnish a physical
interpretation of the particular combination of parameters $\Gamma$. \ First,
note that the kinematic limit $\Gamma\rightarrow0$ can be approached in
several ways, either by decreasing $D$ or by increasing $\left\vert
v-c\right\vert .$ \ The limits $c\rightarrow\pm\infty$ (infinite phase speed)
both correspond to oscillations which are uniform in space. \ Diffusion can
have no effect on a system with no spatial gradients, and so it is
understandable that the effective diffusion vanishes in this limit. \ On the
other hand, \ as $c$ approaches $v$ from either direction $\Gamma$ diverges.
\ This divergence can be understood qualitatively if one considers the
diffusionless limit as a \textquotedblleft zeroth order\textquotedblright%
\ approximation to the phase dynamics. \ In this limit, $\Lambda=\Lambda(0)$
is a constant and the physical wavelength $2\pi/k$ shrinks to zero as
$c\rightarrow v$. \ If a nonzero diffusion is then turned on, then it is
natural to expect that it creates more mixing between crests and troughs when
the waves are close together than when they are far apart. \ As the wavelength
shrinks, the effect of diffusion on the waveform increases.

The special case $c=0$ describes stationary waves, which result from a
steady-state boundary condition. \ The combination of parameters controlling
the behavior of stationary waves is the same combination identified in the
linearized analyisis, namely $\left.  \Gamma\right\vert _{c=0}=D/v^{2}.$
\ \ In the case of complex eigenvalues, the condition for the existence of
undamped periodic stationary waves is the one already derived using the
linearized analysis: \ $\Gamma<\alpha/\beta^{2}.$\ \ Above this threshold all
solutions spiral into the fixed point. \ However, since the nature of
solutions to (\ref{travelODEscaled}) is governed only by $\Gamma$, one can
obtain a more general threshold for undamped periodic \emph{travelling} waves
with phase velocity $c.$ \ Such waves are undamped only if
\begin{equation}
\Gamma=\frac{D}{(v-c)^{2}}<\frac{\alpha}{\beta^{2}}. \label{DcThreshold}%
\end{equation}
For given values of $D$ and of the kinetic control parameters that determine
$\alpha$ and $\beta$, \ there is an interval of phase velocities, surrounding
the flow velocity,\ within which travelling waves do not propagate. For the
case of an unstable node or $\beta=0$, on the other hand, the threshold
(\ref{DcThreshold}) diverges and so it appears that there is no such excluded
band of phase velocities. \ We will further examine this issue below.

\subsection{Numerical solutions and qualitative features of nonlinear
waveforms}

The equation (\ref{travelODEscaled}) \ is useful for several reasons. \ First,
\ it shows that the essential features of travelling and stationary waves (the
amplitude and the waveform) depend on a single combination of transport
parameters, $\Gamma$ , and therefore a family of different waves are described
by a single universal function. \ Second, as a general rule, ODE's can be
solved with less computational effort than PDE's, and (\ref{travelODEscaled})
\ allows one to derive wave solutions without solving the full PDE
(\ref{RDAeq}). \ \ Solution methods are described in Appendix
B.\ (\ref{travelODEscaled}) is solved on a finite interval with boundary
conditions, and the solution will in general include transients near the
boundaries. \ In the case $c=0$, these may represent actual transients of the
stationary wave solution near the physical boundaries of the reactor. \ On the
other hand, if we wish to describe the asymptotic behavior of travelling wave
solutions with $c\neq0$ the transients should be ignored.

\begin{figure}
[ptb]
\begin{center}
\includegraphics[
height=4.4624in,
width=5.3056in
]%
{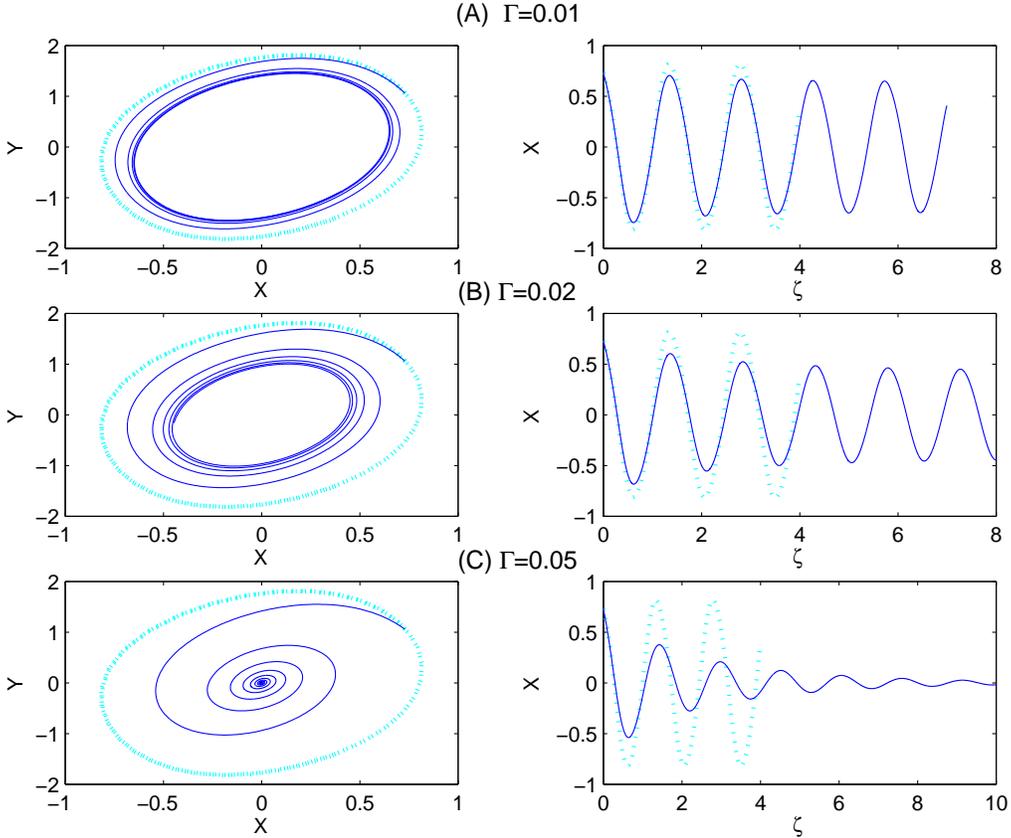}%
\caption{Numerical solutions of equation (\ref{travelODEscaled}) for the FN
system with $e=2$. \ The left column shows $Y(\zeta)$ vs. $X(\zeta)$, the
right shows $X(\zeta)$ vs. $\zeta$. The dotted and lighter-colored curves show
solutions in the kinematic limit (or $\Gamma=0$). \ The boundary conditions
are fixed at a point on the local limit cycle at one end, free at the other
end. For $\Gamma<0.28$ ((A) and (B)) \ there is a periodic quasi-sinusoidal
waveform. \ For $\Gamma>0.28$ (C) the solution spirals into the origin. \ The
period changes little as $\Gamma$ increases. \ }%
\label{qsBVP}%
\end{center}
\end{figure}

\begin{figure}
[ptb]
\begin{center}
\includegraphics[
height=4.4642in,
width=5.3878in
]%
{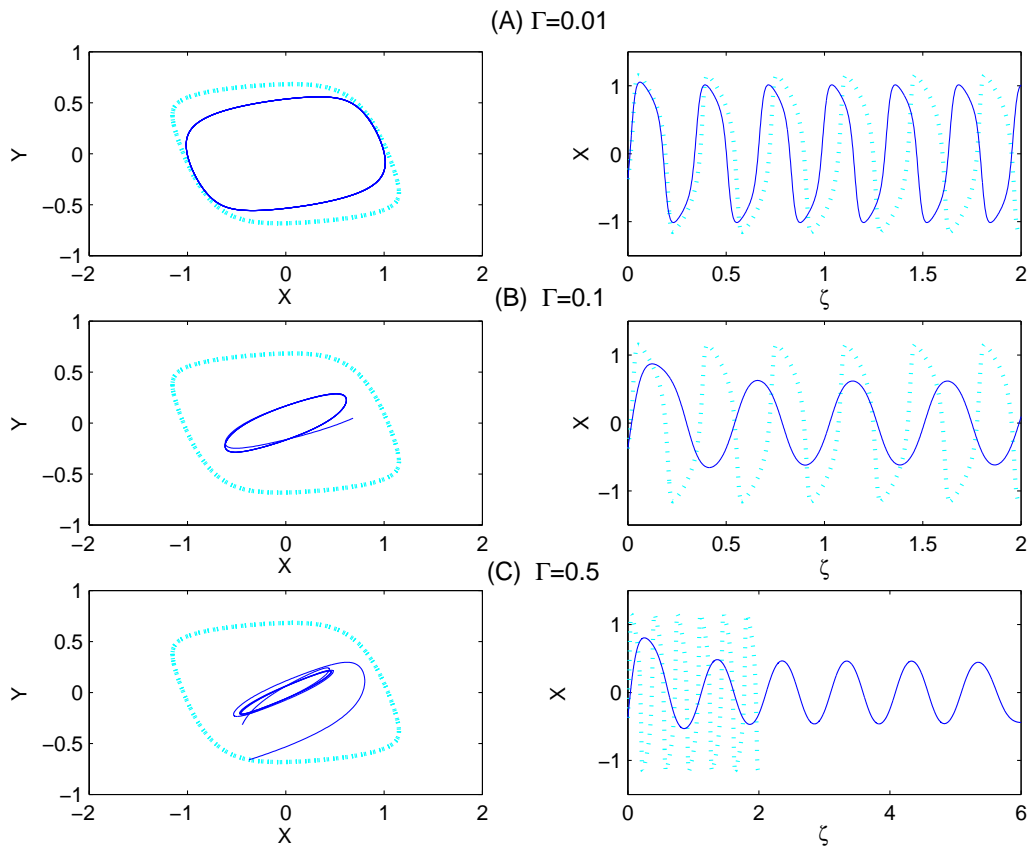}%
\caption{Solutions of equation (\ref{travelODEscaled}) as in figure
\ref{qsBVP} but for $e=50$. \ As $\Gamma$ increases, the period lengthens
compared to the kinematic limit and the waveforms trace narrower loops in
phase space, but they do not spiral into the origin at any finite $\Gamma$.
\ }%
\label{relaxbvp}%
\end{center}
\end{figure}

Some examples of numerical solutions for the FN system are shown in figures
\ref{qsBVP} and \ref{relaxbvp}. \ As expected, increasing $\Gamma$ has the
general effect of increasing the deviations from the kinematic limit.
\ However, the behaviour differs qualitatively between the quasi-sinusoidal
and the relaxation oscillation cases. \ In the former case, \ the phase space
orbit remains approximately elliptical. \ Its period remains approximately
constant while its amplitude shrinks uniformly until, at the critical
threshold $\Gamma=\alpha/\beta^{2}$, it vanishes into the fixed point. \ In
the relaxation oscillation case, on the other hand, the limit cycle does not
shrink to a point. Instead, as $\Gamma$ increases, the period $\Lambda$
increases apparently without bound, and the limit cycle remains elongated in
the more unstable eigendirection while narrowing in the transverse direction.%

%

\subsection{\textquotedblleft Non-linear dispersion relation\textquotedblright%
}

By finding $\Lambda(\Gamma)$ numerically for a range of values of values of
$\Gamma$ and then using the relation (\ref{biglambda}), one can find the
\textit{non-linear dispersion relation} between the frequency $\omega$ (which
can be set by the forcing frequency of a perturbation at the inflow) and the
phase velocity $c$ of large-amplitude travelling waves at given values of the
transport parameters $D,v$. In the quasi-sinusoidal dynamical regime near the
Hopf bifurcation, \ $\Lambda$ is nearly constant and approximately equal to
the small-amplitude oscillation period $2\pi/\beta$. \ \ The $\omega$-$c$
relation is then approximately the same as that predicted in the kinematic
limit, except that it is truncated at the cutoff frequencies $\beta\pm
\sqrt{\alpha v^{2}/D\text{ }}$ where the amplitude vanishes (see figure
\ref{nldispplus}B). \ Outside of this interval of frequencies (which is the
same as the range of amplified frequencies in the linearized theory) there are
only evanescent waves. \ Because of the inverse relation between $\omega$ and
$v-c$, there is, as mentioned above, a range of excluded phase velocities near
$v$ for which undamped travelling waves do not exist. \ 

\begin{figure}
[ptb]
\begin{center}
\includegraphics[
height=1.9899in,
width=2.6991in
]%
{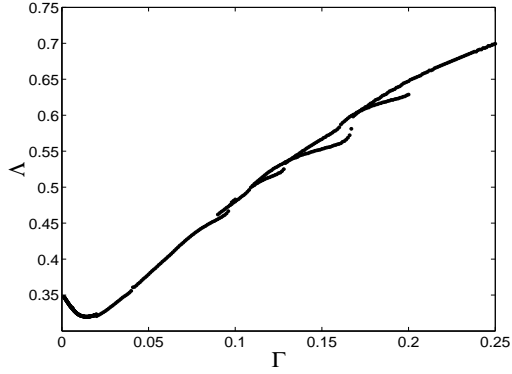}%
\caption{Scaled wavelength $\Lambda$ as a function of $\Gamma$ obtained from
numerical solutions of the one-dimensional equation. \ $\Lambda$ appears to
increase approximately linearly for large $\Gamma$. \ The numerical results
become more uncertain at longer wavelengths due to the finite interval of the
solution (see Appendix B). }%
\label{lambda}%
\end{center}
\end{figure}

\begin{figure}
[ptbptb]
\begin{center}
\includegraphics[
height=2.9931in,
width=5.8548in
]%
{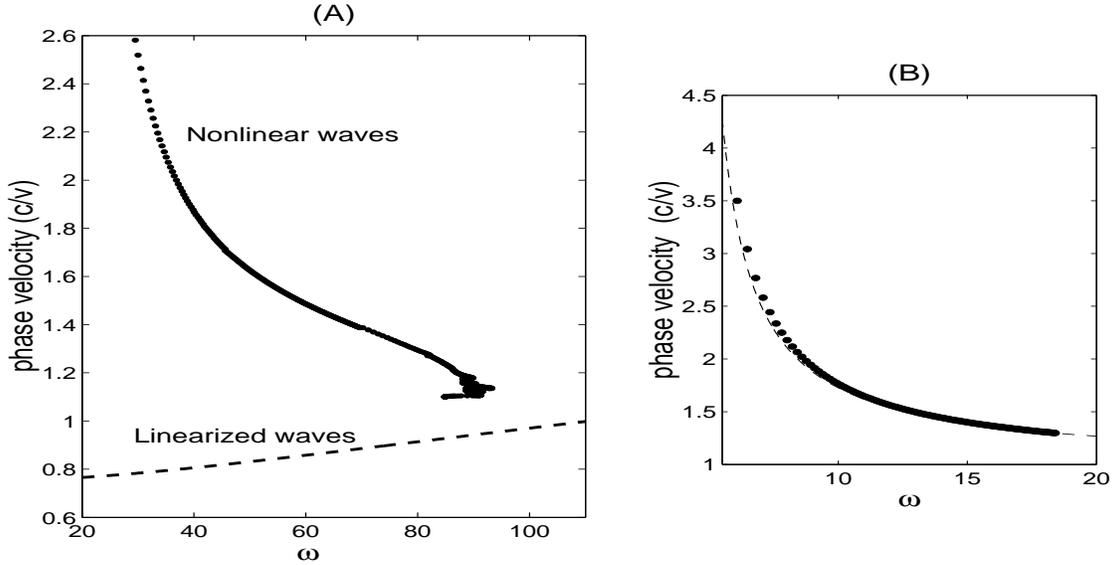}%
\caption{(A) Phase velocity vs. frequency $\omega$ for the case $e=50$, $v=1$,
$D=0.003$ based on the numerical data of figure \ref{lambda}. \ There is
evidently a maximum frequency. \ Perturbations near this frequency generate
waves with a speed close to the flow velocity. \ \ The phase velocity for
linearized small-amplitude waves along the most unstable eigendirection is
shown for comparison. \ Note that the cutoff frequency for small-amplitude
waves occurs precisely when their phase velocity reaches 1, as can be deduced
from the dispersion relation. \ The maximum frequency for large-amplitude
waves (approximately $90$) appears to be lower than the linear cutoff
frequency of $110$. \ In the gap between these two frequencies, the linearized
analysis predicts a growing mode but eq. (\ref{travelODEscaled}) gives no
solution with the correct frequency. \ (B) A similar plot for $e=2,$ $v=1,$
$D=0.0025$. \ In this case the phase velocity for nonlinear waves is very
close to the linearized prediction, and becomes closer as the cutoff frequency
of $18.36$ is approached. \ \ }%
\label{nldispplus}%
\end{center}
\end{figure}

In the relaxation regime, on the other hand, $\Lambda$ varies quite strongly
with $\Gamma.$ \ In fact, numerical results suggest that for asymptotically
large $\Gamma$ it increases approximately linearly (see figure \ref{lambda}).
\ This means that the frequency does \emph{not} become infinite as
$c\rightarrow v$ but that instead there is a maximum.\ \ Such a maximum is
seen in figure \ref{nldispplus}A, which shows frequency versus phase speed for
the case $e=50$, $D=0.003$, $v=1$, based on the numerical data for
$\Lambda(\Gamma)$. \ This relation is radically different from the one for
small-amplitude, linearized waves. \ The maximum frequency appears to be lower
than the cutoff frequency obtained from the linear stability analysis. \ There
is thus a range of frequencies for which the linear theory predicts a growing
mode, yet there are no large-amplitude solutions described by
(\ref{travelODEscaled}) corresponding to these frequncies. \ Numerical results
described below suggest that within this frequency gap the small-amplitude
waves are subject to a secondary instability and do not penetrate far beyond
the inflow boundary. \ \ \

\subsection{Evolution and asymptotic waveforms of growing modes}

\begin{figure}
[ptb]
\begin{center}
\includegraphics[
height=4.4555in,
width=5.9387in
]%
{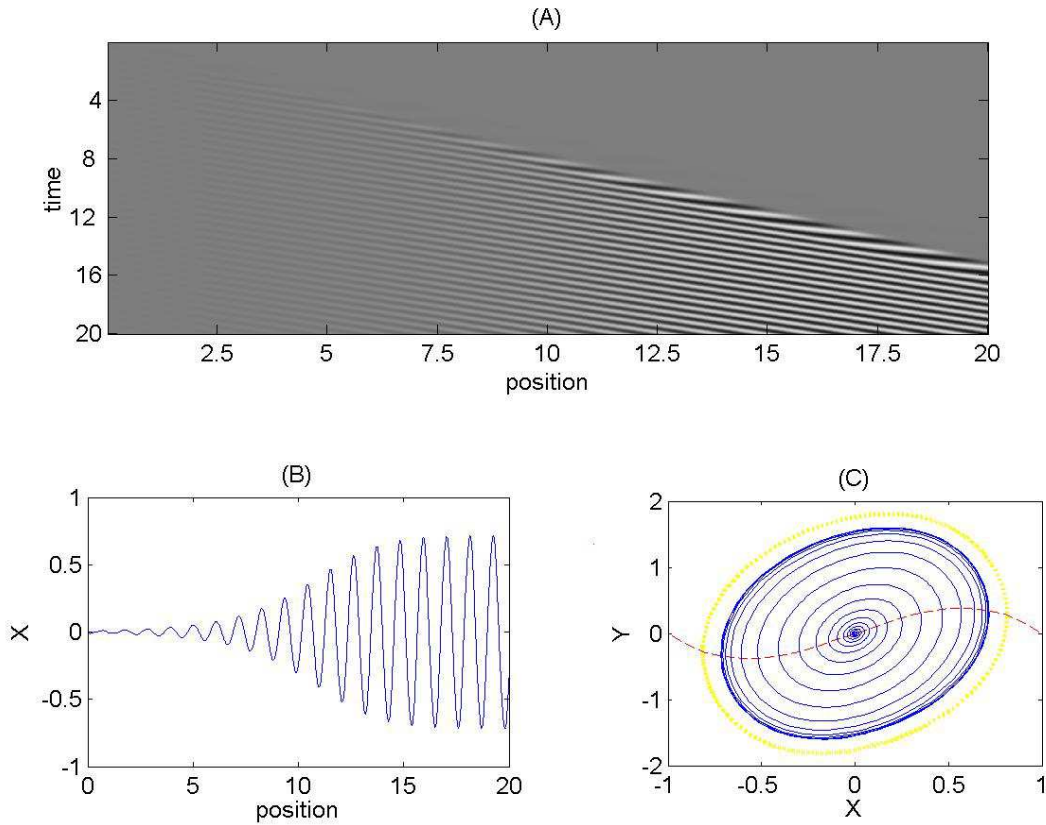}%
\caption{Growing travelling waves for $e=2$, $v=1$, $D=0.0025$ generated by a
sinusoidal boundary perturbation with $\omega=15$. (A) $X(x,t)$ represented as
a gray level. \ (B) $X(x)$ at $t=20$. \ (C) $Y(x)$ vs. $X(x)$ at $t=20$. \ The
limit cycle of the underlying oscillator is shown as a lighter curve, and the
cubic nullcline (thin dashed line) is also shown for reference. \ The actual
waveform has a slightly smaller amplitude than the limit cycle of the
well-stirred system. \ The saturated waves have a phase velocity of 1.65,
giving $\Gamma\approx0.006$.}%
\label{growtravellow}%
\end{center}
\end{figure}

\begin{figure}
[ptbptb]
\begin{center}
\includegraphics[
height=4.4555in,
width=5.7216in
]%
{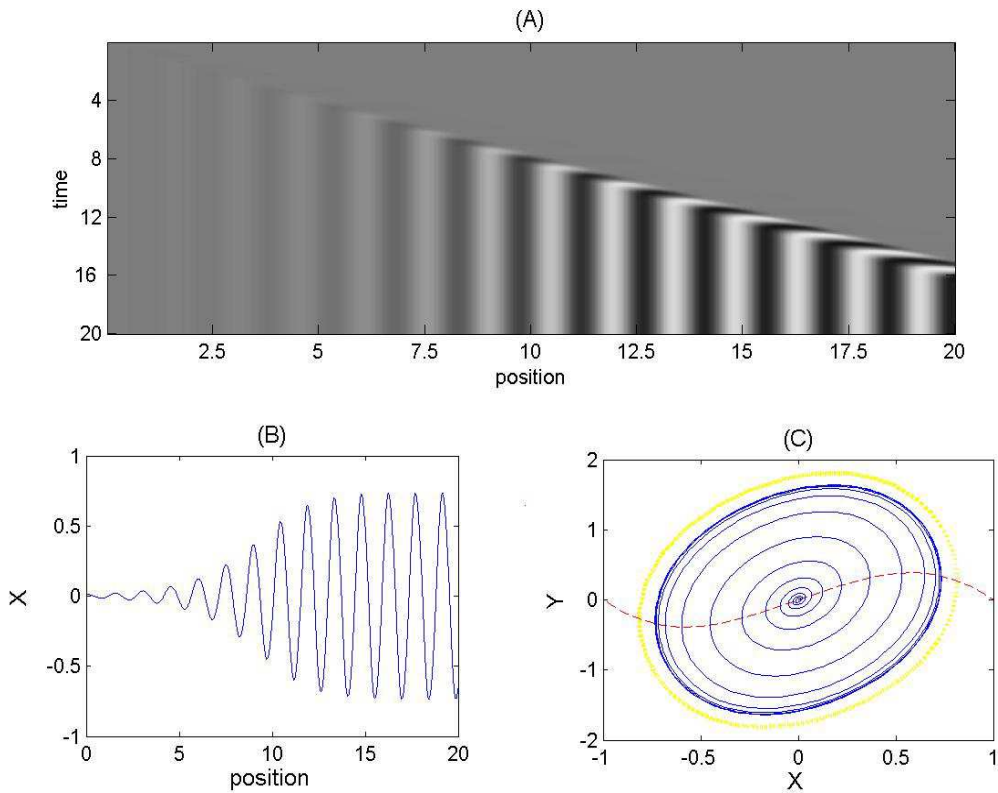}%
\caption{Stationary waves in the FN system with $e=10,$ $v=1$ and $D=0.006$
have the same value of $\mathcal{\Gamma}$ as the travelling waves in figure
\ \ref{growtravellow}. \ Their spatial waveform is similar in shape and
amplitude and deviates from the kinematic limit (lighter curve) by the same
amount. \ }%
\label{statwavelow}%
\end{center}
\end{figure}

In this section we present a few numerical solutions of the PDE (\ref{RDAeq})
with sinusoidal boundary forcing $(X(0,t),Y(0,t))=(a\cos\omega t,a\sin\omega
t)$ where $a$ is a small perturbation amplitude ($a=0.05$ for most examples).
These examples illustrate the principles discussed above, including the role
of the reduced parameter $\Gamma$ and the growth of small perturbations into
nonlinear waves that obey (\ref{travelODEscaled}). \ First, we illustrate the
\textquotedblleft equivalence\textquotedblright\ of waves with the same value
of $\Gamma$. \ Figures \ref{growtravellow} and \ref{statwavelow} are
space-time diagrams of the FN system at $e=2$. \ The diffusion constants are
different, and in one case $\omega=15$ while $\omega=0$ in the other. \ But in
both cases $\Gamma\approx0.006$ and so the functions $\mathbf{U}(x)$ for fixed
$t$ have approximately the same shape and saturate at almost the same
amplitude (slightly below the kinematic limit). \ %


\begin{figure}
[ptb]
\begin{center}
\includegraphics[
height=4.2834in,
width=5.8133in
]%
{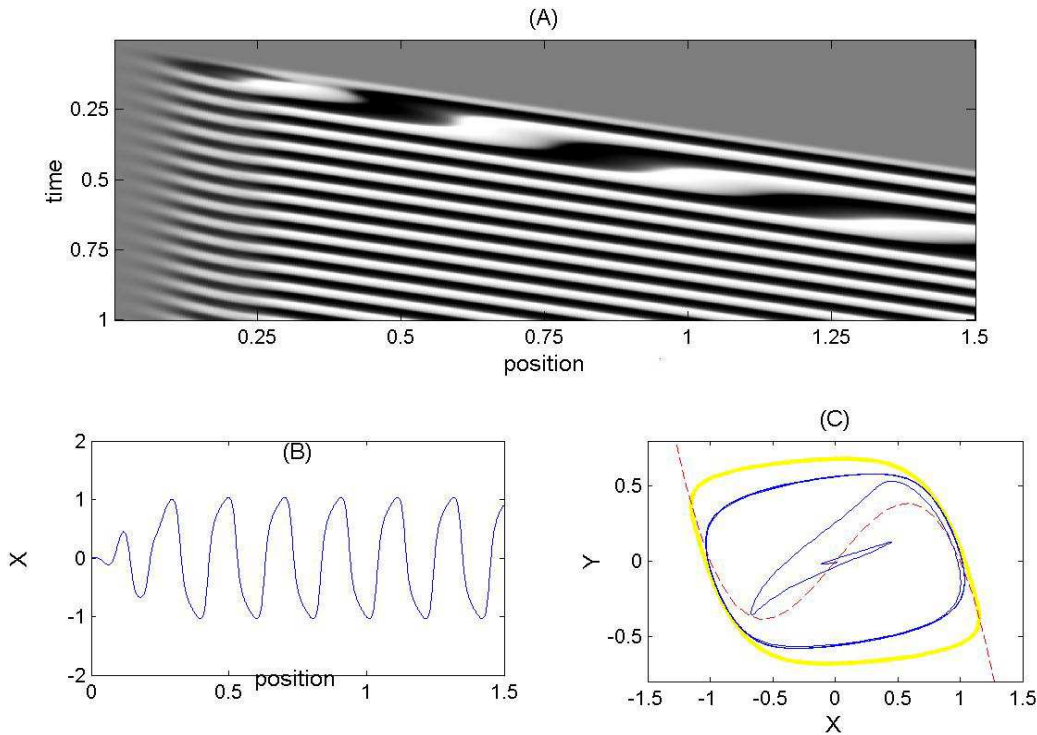}%
\caption{Waves generated by a sinusoidal perturbation with $\omega=50$ in the
FN system with $e=50$, $v=1$ and $D=0.003.$ \ The initial transient response
to the switching on of the perturbation at $t=0$ is followed by a steady
travelling wave with phase velocity $c\approx1.6$, a value consistent with the
$\omega$-$c$ relation plotted in figure \ref{nldispplus}. \ The corresponding
value of $\Gamma$ is $0.008$\ (C) shows deviations from the kinematic limit---
the shape of the local orbit is shown as the broken line. \ Compare (b) and
(c) to figure \ref{relaxbvp}a. \ \ }%
\label{om50}%
\end{center}
\end{figure}

\begin{figure}
[ptbptb]
\begin{center}
\includegraphics[
height=4.4555in,
width=5.751in
]%
{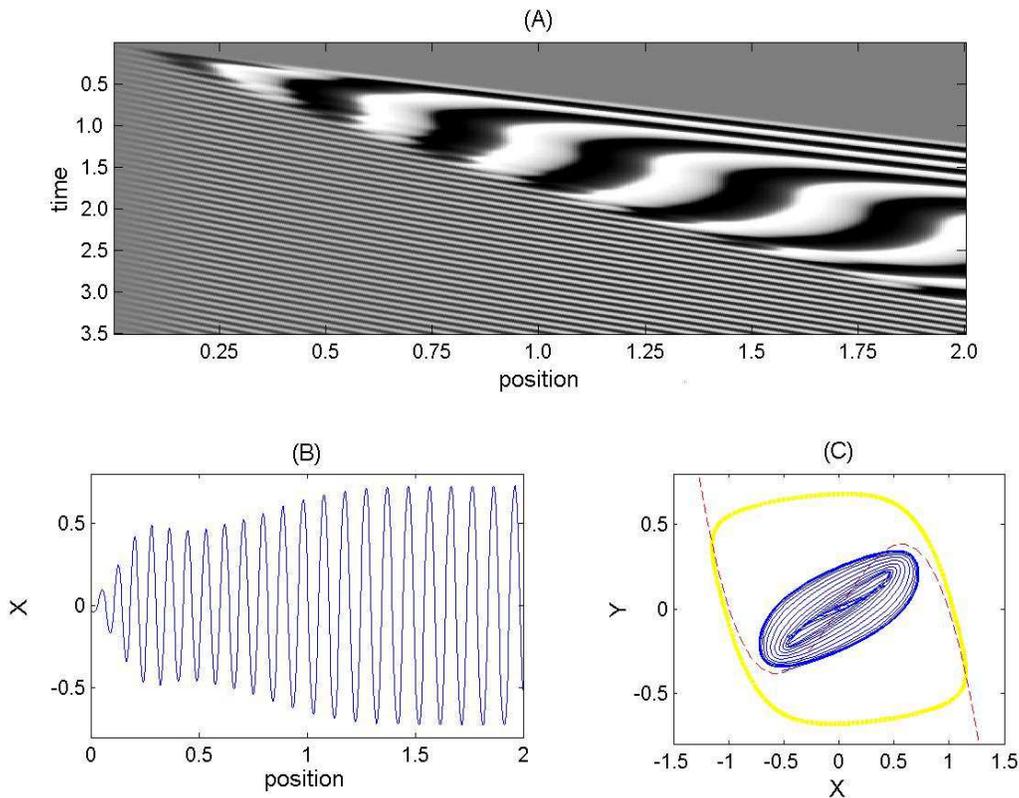}%
\caption{Waves generated by a perturbation with $\omega=80$ show a larger
deviation from the kinematic limit. \ The waves have $c=1.27$ and
$\Gamma=0.04$. \ Compare (b) and (c) with figure \ref{relaxbvp}b.}%
\label{om80}%
\end{center}
\end{figure}

\begin{figure}
[ptbptbptb]
\begin{center}
\includegraphics[
height=4.4538in,
width=5.8358in
]%
{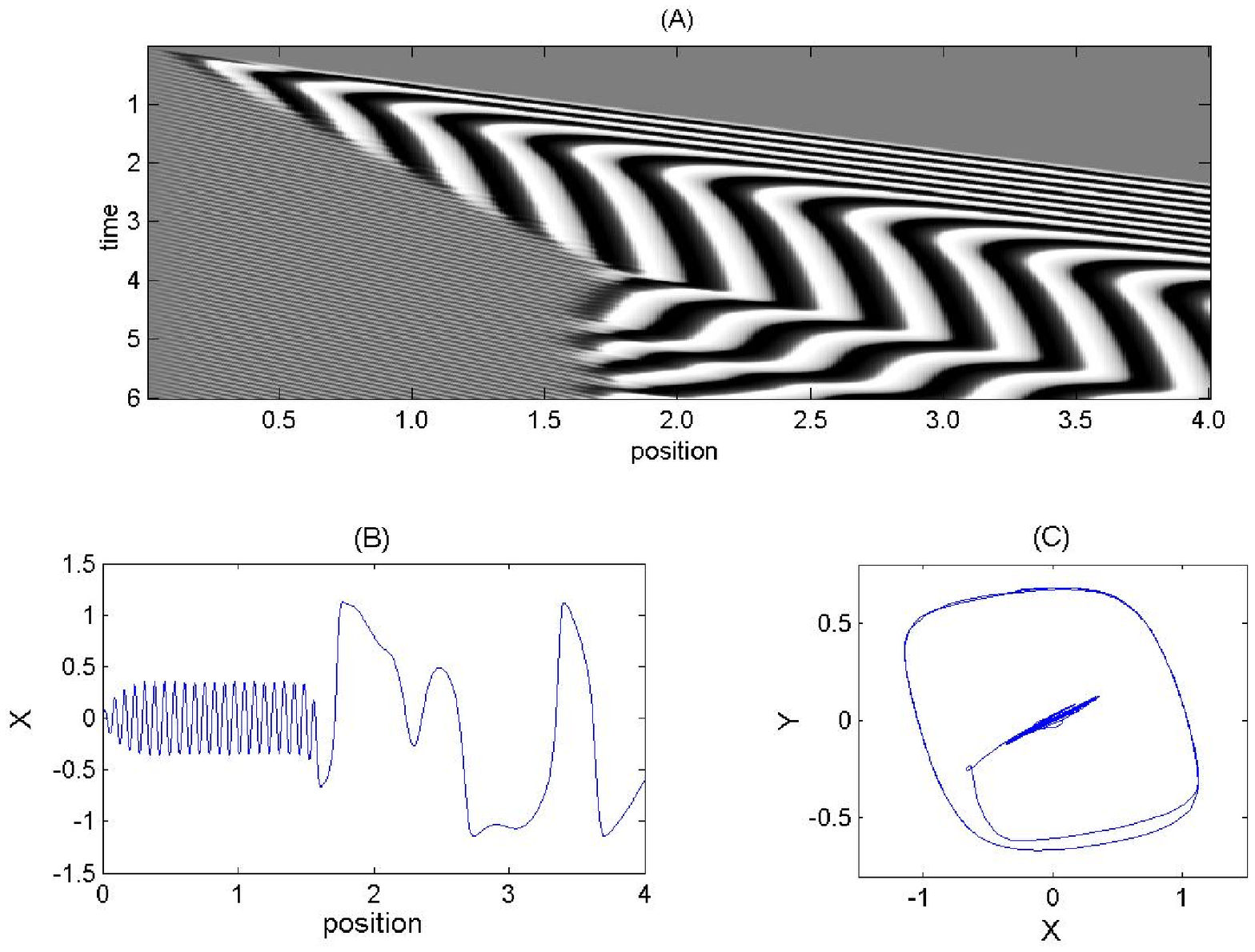}%
\caption{Boundary perturbation frequency $\omega=85$. \ \ A high-frequency
travelling wave, oscillating almost entirely along the most unstable
eigendirection, penetrates a limited distance into the medium before giving
way to a pattern of much longer, irregularly moving waves. \ }%
\label{om85sudden}%
\end{center}
\end{figure}

Next, we examine numerical results for $e=50$, $v=1$ and $D=0.003$, where the
unstable fixed point is a node with two positive real eigenvalues. \ The
predicted $\omega$-$c$ relation for these parameter values is shown in figure
\ref{nldispplus}A. \ The boundary perturbation has components of equal size
along both eigenvectors, but the component along the more unstable eigenvector
of course grows faster. \ Figure \ref{om50} and \ref{om80} show waves
generated by perturbations with $\omega=50$ and $80$, respectively. \ In both
cases, the perturbation grows to a nonlinear travelling wave that deviates
significantly from the kinematic limit (with more pronounced deviations in the
$\omega=80$ case.) \ The waveforms resemble those of figure \ref{relaxbvp}.
\ Figure \ref{om85sudden} shows the results of a perturbation with $\omega=85$
which is near the maximum frequency for nonlinear waves in figure
\ref{nldispplus}. \ In this case, \ the perturbation initially grows along the
more unstable eigendirection, but it penetrates only a small distance into the
medium before breaking up. \ Similar results were found for perturbations
between $\omega\approx85$ and the linear cutoff frequency of $\omega
\approx110$ (For $\omega>110$ the perturbations are immediately damped.)
\ Evidently the waves within this frequency range are subject to a secondary
instability. \ The patterns which arise after the high-frequency waves break
up appear similar to the pulsating waves observed in refs. \cite{Kaern2000}
and \cite{Kaern2002}. \ The behavior of perturbations near the cutoff
frequency in the case of an unstable node warrants further study. \ \ \ \ \ \

\section{Conclusions and discussion}

We have attempted to give a general framework for understanding the behavior
of flow-distributed waves in one-dimensional open flows of oscillatory media
without differential transport, aiming at generic results. \ From the point of
view of linear stability analysis, we examined some changes that occur
generically when an unstable fixed point changes from a focus to a node, \ as
typically occurs at sufficient distance from a Hopf bifurcation. \ We showed
that when the fixed point is a node, the first mode to become absolutely
unstable is a stationary wave mode. In this case, unlike examples previously
noted, a stationary wave can be established by a perturbation of finite
duration in time, at least if the system is not far above the absolute
instability threshold. \ We derived general expressions ( eq. (\ref{amprange}%
)) for the band of frequencies that result in growing waves. \ The center of
this band occurs at zero frequency if the unstable fixed point is a node, and
at a non-zero frequency if it is a focus. \ The expression for the growing
frequency band allows easy calculation of the threshold for the extinction of
stationary waves. \ We then showed that the nonlinear behavior of periodic
travelling or stationary waves reduces to a one-dimensional ODE
(\ref{travelODEscaled}) governed by the single parameter $\Gamma=D/(v-c)^{2}$,
which has a physical interpretation as the strength of diffusive mixing
between peaks and troughs of a particular wave. \ The ODE can be solved
numerically to derive the waveforms and obtain a relation between the
frequency of driving at the boundary and the wavenumber and/or phase velocity
of the waves generated by the perturbation. \ We examined deviations of
waveforms from the kinematic limit, noting qualitative differences between the
quasi-harmonic and relaxation cases. \ \ 

We illustrated our formalism by applying it to the FitzHugh-Nagumo toy model,
but the tools we developed here can be applied to other kinetic models. \ In
particular, we plan to use the current results to analyze FDO patterns of
complex and chaotic oscillators in a future publication. \ The reduced
one-dimensional equation can be applied to systems with multiple fixed points,
subcritical Hopf bifurcations, Canards and bistable behavior, or other
situations in which the linearized analysis gives only limited insight.

Some other questions have been left open. \ The behavior of travelling waves
near the cutoff frequencies in the case of an unstable node may be a fruitful
subject for further study. \ More generally, we have only hinted at the
possibility of secondary instabilities that may affect FDO waves. \ Also, in
this work we have only considered regular travelling waves, not the pulsating
waves observed in refs. \cite{Kaern2000} and \cite{Kaern2002}, although some
of our numerical results (see figure \ref{om85sudden}) \ appeared to show
pulsating waves arising from a secondary instability. \ 

\section{Appendix A: \ The FitzHugh-Nagumo Model, quasi-sinusoidal and
relaxation oscillations}

The FitzHugh-Nagumo (FN) model, also known as the van der Pol oscillator, \ is
defined by the following pair of equations\cite{Muratov}\cite{FN1Dawson}%
\cite{FN2Hagberg}\footnote{This version of the model was used in ref.
\cite{Faraday}.}:%
\begin{align}
\frac{dX}{dt}  &  =e(X-X^{3}-Y)\label{FNgeneral}\\
\frac{dY}{dt}  &  =-Y+aX+b.\nonumber
\end{align}
It is not meant to be a realistic model of any chemical system, \ since its
state variables include negative values, but it serves as a useful toy model
that exhibits many generic features seen in real chemical systems, including
bistability, excitability and ocillations of quasi-sinusoidal as well as
relaxational character.

\begin{figure}
[ptbh]
\begin{center}
\includegraphics[
height=2.0176in,
width=5.1318in
]%
{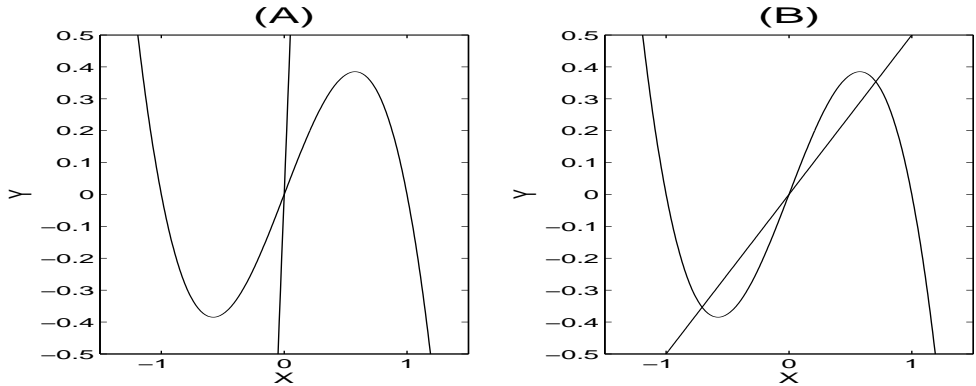}%
\caption{Nullclines for the FHN model for $b=0$ and (A) $a>1$; (B) \ $a<1.$ In
the latter case, there are three fixed points.}%
\label{nullclinepic}%
\end{center}
\end{figure}

The nullclines (Fig. \ref{nullclinepic}) are a cubic and a straight line. $e$
is the ratio of time scales of motion along $X$ and $Y$, $a$ is the slope of
the $Y$ nullcline and $b$ is its intercept with the $Y$-axis. relaxation
oscillations occur when $e$ is large. \ The number and location of the fixed
points depends on the \ $Y$ nullcline, i.e. on values of $a$ and $b$. There
may be either one fixed point or three, as shown in fig. \ref{nullclinepic}.
\ A fixed point $(X_{\ast},Y_{\ast})$ is always stable when it lies outside
the extrema of the cubic, i.e. for $X_{\ast}<-1/\sqrt{3}$ and $X_{\ast
}>1/\sqrt{3}$. Depending on the parameters $(e,a,b)$ a single fixed point may
be unstable and give rise to a limit cycle if $-1/\sqrt{3}<X_{\ast}<1/\sqrt
{3}$. \ \ One case that has often been studied is that of a stiff system with
a large time scale separation $e\gg1$ and a single fixed point. \ In this
case, as $b$ is adjusted so as to move the fixed point closer to the origin,
there is first a Hopf bifurcation at a fixed point near one extremum of the
cubic nullcline, and then a Canard transition\cite{Canard} in which the limit
cycle changes rapidly from a small-amplitude one to a large-amplitude
relaxation oscillation. \ However, we do not consider the Canard transition in
this paper but instead follow reference \cite{Faraday} in setting $b=0$, thus
ensuring a fixed point at the origin. By setting $a=10$ we also ensure that
there is only one fixed point, \ thus excluding bistable or excitable
behavior. \ In the examples we discuss, we vary only $e$. \ \

\begin{figure}
[ptbh]
\begin{center}
\includegraphics[
height=2.3506in,
width=3.9825in
]%
{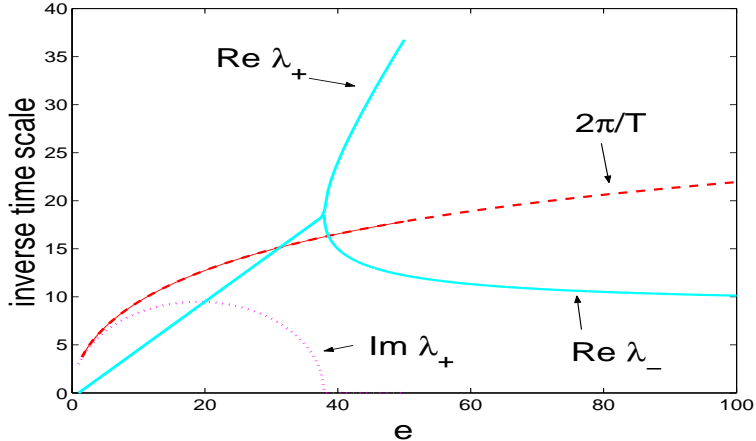}%
\caption{Inverse time scales as functions of $e$ for $a=10$. \ Dotted line:
$\operatorname{Im}(\lambda_{+})$; \ solid lines: \ $\operatorname{Re}%
(\lambda_{\pm})$; \ dashed line: \ $2\pi/T$ for the limit cycle. \ At the
critical value $e_{c}\approx38$, the eigenvalues become real. \ }%
\label{freqplot}%
\end{center}
\end{figure}

\begin{figure}
[ptbhptbh]
\begin{center}
\includegraphics[
height=3.6694in,
width=4.753in
]%
{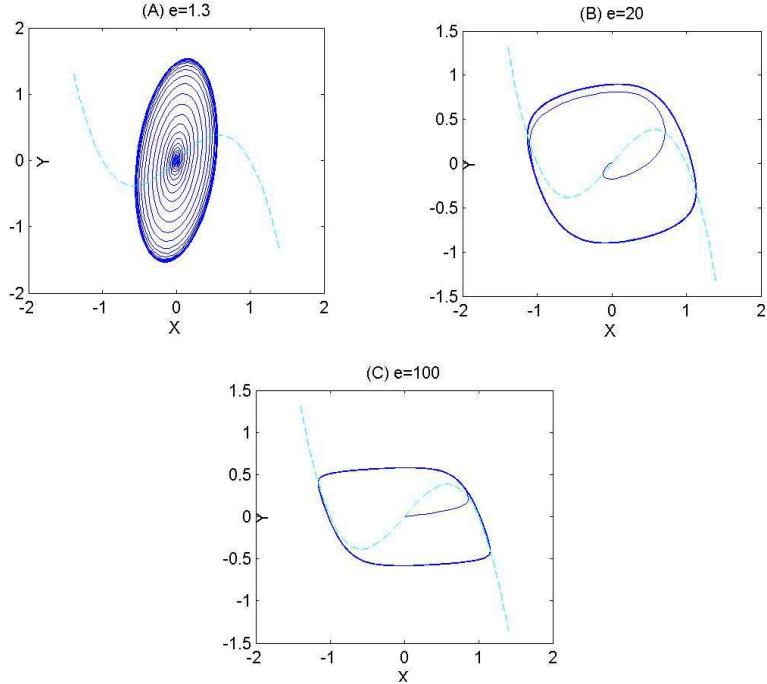}%
\caption{Trajectories of an initial condition close to the origin, for $a=10$
and three different values of $e.$ (A)\ $e=1.3$ is close to the Hopf
bifurcation and the oscillations are quasi-sinusoidal. \ (B) $e=20$ is an
intermediate value, between the Hopf bifurcation and $e_{c}\approx38$. \ (C)
$e=100$ is well above $e_{c}$ and the limit cycle has the character of a
relaxation oscillation. \ \ The change in the shape of the limit cycle is a
gradual one, in spite of the rapid transition in the eigenvalues at the fixed
point.}%
\label{traj3}%
\end{center}
\end{figure}

The Jacobian evaluated at the origin is given by%
\begin{equation}
\left(
\begin{array}
[c]{cc}%
e & -e\\
a & -1
\end{array}
\right)  \label{Jacobian}%
\end{equation}
and its two eigenvalues are given by%
\begin{equation}
\lambda_{\pm}=\frac{e-1}{2}\pm\frac{1}{2}\sqrt{(1+e)^{2}-4ea}.
\label{eigenvalues}%
\end{equation}
The eigenvalues are complex if and only if%
\[
a>\frac{(1+e)^{2}}{4e}=\frac{1}{4e}+\frac{1}{2}+\frac{e}{4}.
\]
Otherwise, both eigenvalues are real, and their signs are the same if $a>1$.
\ If the eigenvalues are complex, then their real part is positive when $e>1.$
\ When $a=10$, the eigenvalues are real and positive for all $e>e_{crit}%
\approx38$. \ Figure \ref{freqplot} shows the real and imaginary parts of the
two eigenvalues $\lambda_{\pm}$ together with the angular frequency
$\omega_{LC}=2\pi/T$ of the stable limit cycle which exists for all $e>1$.
\ The Hopf bifurcation occurs at $e=1$ where $\operatorname{Re}(\lambda_{\pm
})$ becomes positive. \ In the immediate vicinity of the Hopf bifurcation, the
frequency of the limit cycle is identical to the imaginary part
$\operatorname{Im}(\lambda_{+})$. \ At $e_{crit}$, however,
\ $\operatorname{Im}(\lambda_{+})$ vanishes and the eigenvalues become real.
\ They are degenerate at the critical point but quickly become different as
$e$ increases further. \ Roughly speaking, the two real eigenvalues above
$e_{crit}$ correspond to two different inverse time scales: \ a slower one for
motion in the $Y$ direction and a faster one for motion in the $X$ direction.
\ It is this separation of time scales that distinguishes relaxation
oscillations from quasi-sinusoidal ones. \ The frequency of relaxation
oscillations is determined primarily by the \emph{slower} of the two time
scales. \ As $e$ increases, the qualitative character of the oscillations
changes from approximately sinusoidal to relaxation oscillations, as shown in
figure \ref{traj3}. \ Figure \ref{traj3} shows trajectories starting from
points close to the fixed point, therefore it also illustrates visually the
change in character of the fixed point from an unstable focus (trajectories
spiral away from the fixed point) to a node (trajectories leave along the most
unstable eigenvector). \ Note that although there is a sudden change in the
eigenvalues and eigenvectors near the fixed point at $e_{crit}$, the
associated change in the nonlinear limit cycle is gradual. \ \

\section{Appendix B: Numerical solutions of the reduced one-dimensional
equation (\ref{travelODEscaled})}

Here we discuss the solution of the reduced ODE (\ref{travelODEscaled}):%
\[
0=\mathbf{f}(\mathbf{u})-\frac{d\mathbf{u}}{d\zeta^{\prime}}+\Gamma_{c}%
\frac{d^{2}\mathbf{u}}{d\zeta^{\prime2}}.
\]
In the kinematic limit $\Gamma\rightarrow0$, \ this equation reduces to a
first-order equation, identical in form to that of the dynamics of the
well-stirred system. \ Solutions of this first-order equation with general
initial conditions approach the stable limit cycle of the well-stirred system.
\ However, for any nonzero value of $\Gamma$ the equation is second-order and
anti-dissipative\footnote{If the equation is rearranged to isolate the
second-order on one side, the analogy with the equation of motion of a point
particle shows that the \textquotedblleft force\textquotedblright\ contains an
anti-damping term, and the term $-\mathbf{f}(\mathbf{u})$ is also of the
\textquotedblleft wrong\textquotedblright\ sign, tending to push the particle
away from the stable limit cycle of the well-mixed system. \ }, so that the
initial-value problem leads to unbounded solutions for most choices of initial
conditions $\mathbf{u}(0)$ and $\mathbf{u}^{\prime}(0)$. \ To exclude these
unphysical solutions, a boundary condition must be imposed. \ Thus, the
equation should be solved as a boundary-value problem on a finite interval
$0<\zeta<L$. \ Our procedure was to impose a fixed boundary condition on the
left, \ $\mathbf{u}(0)=\mathbf{u}_{in}$ and a free boundary condition on the
right, $\mathbf{u}^{\prime}(0)=0$. \ In the case $c=0$ the boundaries
correspond directly to the the physical boundaries of the plug-flow reactor. \ 

For the examples studied, we found that for moderate values of $\Gamma$ and
for sufficiently large $L$ the solutions of the boundary problem behave
qualitatively like solutions of a first-order initial value problem with an
attractor. \ In other words, \ after some transient behavior at small $\zeta$
which depends strongly on $\mathbf{u}_{in}$, \ the solutions settle either to
a fixed value or to a periodic behavior with an intrinsic period $\Lambda
$,\footnote{A chaotic attractor is also possible. \ We plan to discuss this in
a future publication. \ } which depends on $\Gamma$ but does \emph{not} depend
sensitively on $\mathbf{u}_{in}$ or on $L$. \ The free boundary condition on
the right affects the solution only in a small interval near the right
boundary, i.e., \ the boundary can be moved to a larger value of $\zeta$
without changing the solution on most of the interval $0<\zeta<L$. \ In the
$c=0$ case, the entire solution, including the boundary transients, is
physically meaningful as part of a stationary wave pattern in the reactor.
\ For $c\neq0$, boundary conditions at fixed $\zeta$ values are not directly
equivalent to boundary conditions at fixed $x$, \ but the attractors reached
by the solutions can be interpreted as the asymptotic shapes of travelling
waves in the medium. \ 

In order to obtain numerical solutions of the boundary value problem, we used
a collocation algorithm included in the Matlab software package.\cite{Matlab}
\ This algorithm requires an initial guess for the solution which is then
adjusted to satisfy the differential equations and the boundary conditions
within a specified tolerance. \ For long solution intervals (large multiples
of $\Lambda$) the algorithm may fail to converge unless the initial guess is
close to the actual solution. \ We used two procedures for iteratively
obtaining a solution:

\begin{enumerate}
\item Use a solution of the initial-value problem of the first-order,
$\Gamma=0$ system as an initial guess for a relatively small value of $\Gamma
$. \ Then increase $\Gamma$ iteratively to the desired value, using each
solution as the guess for the next value of $\Gamma$. \ This procedure was
used, for example, to generate the solutions at a range of values of $\Gamma$
in figure \ref{lambda}.

\item Sometimes it is more convenient to approximate the solution with
piecewise solutions on a series of overlapping intervals. \ The procedure is
as follows: \ First, \ solve the boundary value problem with the desired
boundary condition $\mathbf{u}_{in}$ at $\zeta=0$ and free boundary conditions
at a relatively small $L$ which is neither too much larger nor too much
shorter than $\Lambda$. \ Obtain a solution $\mathbf{u}_{1}(\zeta)$ on that
interval. \ Then evaluate that solution at $\zeta=L/2$ and use $\mathbf{u}%
_{1}(L/2)$ as the boundary condition for a new solution on the interval
$L/2<\zeta<3L/2$. \ Continue this procedure on a series of overlapping
intervals. \ If $L$ is not too large, then the initial trial solutions need
not be close to the final ones, and if $L$ is not too small they are not
sensitive to the free boundary condition at the right, so that the overlapping
solutions should be approximately the same except very near the boundaries.
\ Stitched together, the piecewise solutions approximate a solution on a
longer interval. \ 
\end{enumerate}

Large values of $\Gamma$ (where large means significantly larger than
$1/4\alpha$, the threshold of absolute instability in the $c=0$ case) often
presented computational challenges because the solutions become more sensitive
to the right boundary condition, and large intervals were needed in order for
an attractor to appear. \ This is the source of some of the numerical jitter
in the data in figure \ref{lambda}.

\bigskip

\textbf{Acknowledgment:} This work was supported by the NSERC of Canada.

\end{document}